%% file: article.tex
\pdfoutput=1

\documentclass[letterpaper, 10 pt, conference]{ieeeconf}  

\IEEEoverridecommandlockouts                              
\usepackage{enumerate}
\usepackage{url}
\usepackage{graphicx}
\usepackage[table]{xcolor}
\usepackage{booktabs}
\usepackage{verbatim}
\makeatletter
\newcommand{\verbatimfont}[1]{\def\verbatim@font{#1}}%
\makeatother
\verbatimfont{\ttfamily\small}
\usepackage{multirow}
\usepackage{array}
\usepackage{amsmath}
\usepackage{centernot}

\usepackage{tabu}
\usepackage{amssymb}
\usepackage{graphics}
\usepackage{mathrsfs}
\usepackage{cite}
\usepackage[mathscr]{euscript}
\usepackage{subcaption}
\include{philipMacros}

\newcommand{\es}{{\rm ES}}
\newcommand{\opt}{{\rm opt}}
\allowdisplaybreaks

\newtheorem{theorem}{Theorem}

\newtheorem{definition}{Definition}

\title{The Cost of Denied Observation in Multiagent Submodular Optimization}
\author{{David Grimsman, Joshua H. Seaton, Jason R. Marden, and Philip N. Brown}
\thanks{This work is supported by Colorado State Bill 18-086, NSF grant \#ECCS-2013779, Air Force grant FA95550-20-1-0054 and ONR grant N00014-20-1-2359.}
\thanks{D. Grimsman (corresponding author) and J. R. Marden are with the Department of Electrical and Computer Engineering, University of California, Santa Barbara, CA, {\texttt{\{davidgrimsman,jrmarden\}@ece.ucsb.edu}}.}
\thanks{J. Seaton and P. N. Brown are with the Department of Computer Science, University of Colorado Colorado Springs, {\texttt{\{jseaton,philip.brown\}@uccs.edu}}.}
\thanks{© 2020 IEEE. Personal use of this material is permitted. Permission from IEEE must be
	obtained for all other uses, in any current or future media, including
	reprinting/republishing this material for advertising or promotional purposes, creating new
	collective works, for resale or redistribution to servers or lists, or reuse of any copyrighted
	component of this work in other works.}  
}

\begin{document}

\maketitle


\begin{abstract}
A popular formalism for multiagent control applies tools from game theory, casting a multiagent decision problem as a cooperation-style game in which individual agents make local choices to optimize their own local utility functions in response to the observable choices made by other agents.
When the system-level objective is submodular maximization, it is known that if every agent can observe the action choice of all other agents, then all Nash equilibria of a large class of resulting games are within a factor of $2$ of optimal; that is, the price of anarchy is $1/2$.
However, little is known if agents cannot observe the action choices of other relevant agents.
To study this, we extend the standard game-theoretic model to one in which a subset of agents either become \emph{blind} (unable to observe others' choices) or \emph{isolated} (blind, and also invisible to other agents), and we prove exact expressions for the price of anarchy as a function of the number of compromised agents.
When $k$ agents are compromised (in any combination of blind or isolated), we show that the price of anarchy for a large class of utility functions is exactly $1/(2+k)$.
We then show that if agents use marginal-cost utility functions and at least $1$ of the compromised agents is blind (rather than isolated), the price of anarchy improves to $1/(1+k)$.
We also provide simulation results demonstrating the effects of these observation denials in a dynamic setting.
\end{abstract}

\section{Introduction}

Game-theoretic design methods for distributed control of multiagent systems have received considerable attention in recent years, with applications ranging from distributed power generation, swarming of autonomous vehicles, network routing, smart grids, and more~\cite{Saad2012,Xu2012,Marden2017,Kordonis2019,Ferguson2019}.
One common game-theoretic approach involves modeling the agents in a multiagent system as players in a cooperation-style game, designing utility functions for these players and programming the players with a distributed algorithm with which to individually optimize their utility functions~\cite{Gopalakrishnan2010,Marden2012}.
Provided that the designed utility functions and algorithms are selected appropriately, quality and convergence guarantees can then be inherited from the broader game theory literature~\cite{Foster1990,Blume1993,Papadimitriou2001,Alos-Ferrer2010,Marden2013,Papadimitriou2016}.
One of the strengths of this approach is its modularity, as agent utility functions can typically be designed independently from distributed algorithms: the utility functions govern the \emph{quality} of the resulting emergent behavior, and the algorithms govern \emph{dynamics and convergence}.

The above game-theoretic approach has offered a multitude of attractive theoretical guarantees; an early example of this applies to the class of multiagent systems in which the agents are designed to be maximizing a \emph{submodular} function; that is, one that exhibits decreasing marginal returns~\cite{Nemhauser1978}.
Here, the submodularity of the system objective function can be leveraged in combination with a wide variety of utility function designs (yielding the class of so-called \emph{valid utility games}) to ensure that all Nash equilibria are within a factor of 2 of optimal; i.e., the \emph{price of anarchy} is $1/2$~\cite{Vetta2002}.
These types of results have a great deal of synergy with the broader literature on submodular maximization, which itself has a wide application space~\cite{Krause2008,Kempe2003,Barinova2012,clark2011submodular,Lin2011}.


Following from the initial successes of this game-theoretic methodology, recent work has begun to critically investigate the robustness properties of this approach.
For instance, it has been shown that in general settings, slight changes to agent utility functions can lead to dramatic changes in the quality of emergent behavior~\cite{Brown2018g}, and that faulty or misbehaving agents can easily lead stochastic learning dynamics astray~\cite{Jaleel2019a}.
While a comprehensive measure of robustness for such systems remains elusive, positive results exist as well.
In particular, for submodular maximization, it is known that performance guarantees can be quite robust to discrepancies of information availability among the agents when the agents are endowed with the specific \emph{marginal-contribution} utility function~\cite{Gharesifard2018,Grimsman2018a,Brown2019c}.
Specifically, these papers show that the price of anarchy associated with marginal-cost utility functions degrades gracefully as information is denied to agents.
While attractive, these preliminary positive robustness results are limited in scope as they consider only the specific marginal-contribution agent utility function, despite the fact that this is only one possible choice of utility design and is not optimal in all settings~\cite{Gairing2009,Paccagnan2019,Chandan2019}.

Accordingly, this paper initiates a study on the robustness of performance guarantees for the broad class of valid utility games when agent actions are not observable by all agents.
In our model, we study valid utility games in which a set of $k$ agents is compromised either by becoming \emph{blind} (unable to observe the action choices of any other agent, but still observable by others) or becoming \emph{isolated} (unable to observe other agents or be observed by other agents).
Our main result in Theorem~\ref{thm:vug 2+k} states that the price of anarchy when $k$ agents are compromised is $1/(2+k)$, and that this bound is tight for any combination of blind or isolated agents.
This result is significant in at least two dimensions: first, it shows for general valid utility games that, in line with the narrower characterization of earlier work~\cite{Gharesifard2018,Grimsman2018a,Brown2019c}, performance guarantees degrade gracefully as information is denied from agents.
Second, and perhaps more surprisingly, Theorem~\ref{thm:vug 2+k} illustrates that isolation is no worse than blindness.
Intuitively, this suggests that if an agent is blind, it might as well be invisible also.


This raises the question: are blindness and isolation equivalent for all forms of utility functions for the agents?
Our Theorem~\ref{thm:mc good} answers this in the negative, showing that \emph{if} the non-compromised agents are endowed with the marginal-contribution utility function, isolation has a cost: the price of anarchy resulting when $k$ agents are compromised improves if some of the compromised agents are not isolated.
Specifically, if at least $1$ of the $k$ compromised agents is blind (but not isolated), then the price of anarchy improves to $1/(1+k)$.
Thus, Theorem~\ref{thm:mc good} also demonstrates graceful degradation of performance guarantees, and in addition shows that for some utility function designs, blindness can indeed be strictly better than isolation.

\subsection{Model Preliminaries} \label{ssec:model}

A submodular multiagent optimization problem has agent set $N =\left\{1,\ldots,n\right\}$ and a finite set of resources ${\cal R}$.
Each agent $i\in N$ has a set of admissible \emph{actions} given by subsets of resources: $\aaa_i\subseteq 2^{\cal R}$ and denote the \emph{joint action space} by $\aaa:=\aaa_1\times\dots\times\aaa_n$.
For convenience, we assume that every agent can \emph{opt out} by selecting the empty set, or $\emptyset\in\aaa_i$.
The overarching goal is to maximize an objective function $W:\aaa\to\arr$.

For an action profile $a=(a_1,a_2,\ldots,a_n)\in\aaa$ and $J \subseteq N$, let $a_J$ denote the profile of only the actions of agents in $J$. 
Alternatively, one could think of this as the action profile where agents not in $J$ choose to opt out. Likewise, we denote $a_{-i}$ to mean  $a_{N \setminus \{i\}}$ and $a_{-ij}$ to mean $a_{N \setminus \{i, j\}}$. With this notation, we will sometimes write an action profile $a$ as $(a_i,a_{-i})$ or $(a_J, a_{N \setminus J})$. Similarly, we may write $W(a)$ as $W(a_i,a_{-i})$ or as $W(a_J, a_{N \setminus J})$.
Let $\aaa_{-i}=\Pi_{j\neq i}\aaa_j$ denote the set of possible collective actions of all agents other than player $i$.

We denote $R(a) \subseteq \cal R$ as the base set of resources contained in the action profile $a$.
Then objective function $W$ is \emph{submodular} if for all $i\in N$ it holds that
\begin{equation}
    W(a) - W(a_{-i}) \geq W(a_i, a'_{-i}) - W(a'_{-i}),
\end{equation}
for all $a_i$ and for all $a_{-i}, a'_{-i} \in \aaa_{-i}$ such that $R(a_{-i}) \subseteq R(a'_{-i})$.
In other words, we see a decreased marginal contribution for agent $i$ when other agents choose $a_{-i}$ as opposed to $a'_{-i}$.
We also say that $W$ is \emph{nondecreasing} if $W(a) \leq W(a')$ for all $R(a) \subseteq R(a')$, and \emph{normalized} if $W(\emptyset)=0$.
In this work, unless otherwise stated, every objective function $W$ is assumed to be submodular, nondecreasing, and normalized.

\subsection{Utility Design}

A well-studied class of distributed methods for solving submodular optimization problems relies on tools from game theory; at a high level, the agents in the optimization problem are endowed with utility functions which they subsequently attempt to optimize.
Following from~\cite{Marden2014}, a system designer assigns each agent $i$ a utility function $U_i:\aaa\rightarrow\mathbb{R}$ which will guide their decision-making process. 
The designer's goal is to select utility functions that combined with an appropriate learning rule will provably lead the agents to a joint action profile which is of high quality measured by the system-level objective $W$. 
Note that the utility functions $\{U_i\}$ may in general be different from the global objective function $W$.
A multiagent optimization problem of the above form is specified by the tuple $G=\left( N,\aaa,\{U_i\}_{i\in N},W\right)$.
Thus, each problem $G$ is a distributed optimization problem with objective $W$ coupled to a finite game induced by the designed utility functions $\{U_i\}_{i\in N}$.

One utility function that is of note for this work is marginal contribution ($\mc$), wherein each agent maximizes its marginal contribution to the objective function $W$, with respect to the remaining agents. More formally stated:
\begin{equation} \label{eq:mcdef}
    \mc_i (a_i, a_{-i}) := W(a_i, a_{-i}) - W(\emptyset, a_{-i}).
\end{equation}

Of course, this class of problems admits many other utility functions as well.
In this work we consider those which satisfy the \emph{valid utility game} assumptions of~\cite{Vetta2002}:

\begin{definition} \label{def:vug}
A Valid Utility Game (VUG) is a multiagent optimization problem satisfying the following three conditions:
\begin{enumerate}
    \item $W$ is submodular, nondecreasing, and normalized, \label{itm:submod}
    \item $U_i(a_i, a_{-i})\geq W(a_i, a_{-i})-W\left(\emptyset, a_{-i}\right)$ \label{itm:marg}
    \item $\sum_i U_i(a_i, a_{-i})\leq W(a_i, a_{-i})$ \label{itm:sum}
\end{enumerate}
\end{definition}
Note that when $W$ satisfies \ref{itm:submod}), $\mc$ is one possible choice of utility function that satisfies \ref{itm:marg}) and \ref{itm:sum}).

\subsection{Compromised Agents}

We extend the standard model described thus far to include compromised agents of various forms.
We begin with the assumption that $G$ is a VUG, and then some subset of agents is compromised in a way that limits the amount of information they have access to, possibly limiting their decision-making capability. We model this by saying that $U_i$ is replaced by $\tilde{U}_i$ for all agents. We consider three ways in which an agent can be compromised:
\begin{enumerate}
    \item \emph{Blind agents}: a blind agent does not know the actions of any other agents, i.e., if agent $i$ is blind, then $\tilde{U}_i(a) = U_i(a_i, \emptyset)$ and $\tilde{U}_j = U_j$ for $j \neq i$. We denote the set of blind agents as $B\subseteq N$.
    \item \emph{Isolated agents}: an isolated agent is blind, and the other agents are also blind to it. In other words, if agent $i$ is isolated, $\tilde{U}_i(a) = U_i(a_i, \emptyset)$, and $\tilde{U}_j(a) = U_j(a_{-i})$ for $j \neq i$. We denote the set of isolated agents as $I\subseteq N.$
    \item \emph{Disabled agents}: a disabled agent $i$ cannot contribute to system welfare and always selects $a_i=\emptyset$.
    The remaining agents are unaffected: $\tilde{U}_j = U_j$ for $j \neq i$. We denote the set of disabled agents as $D\subseteq N$.%
\footnote{
Note that for the case of blind or isolated agents, if agent $i$ cannot ``see'' the actions of agent $j$, agent $i$ effectively assumes that agent $j$ is disabled.
It should be noted that assuming disability is merely one possibility and that optimal modeling of unobservable agents is an open area of research~\cite{Brown2019c}.
}
\end{enumerate}

\begin{figure}
    \centering
    \begin{subfigure}[b]{0.48\textwidth}
        \centering
        \includegraphics[scale=0.5]{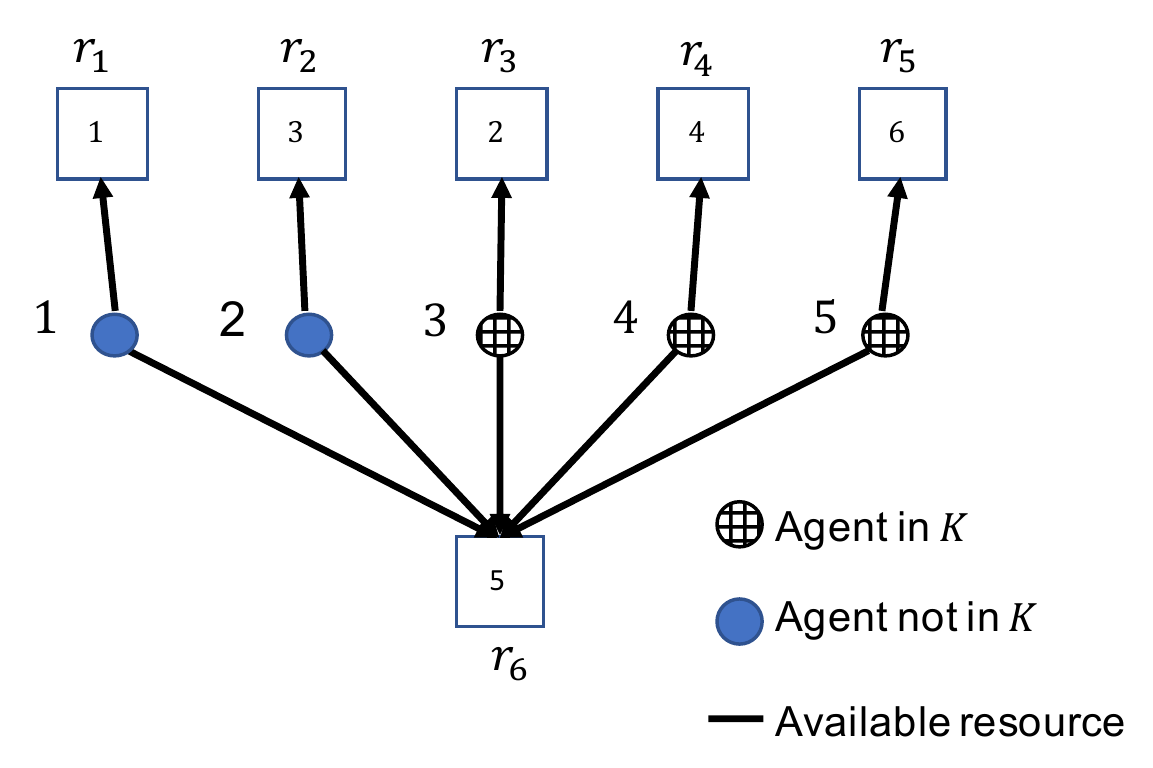}
        \caption{A valid utility game where some agents have been compromised. The agents are represented by circles, with the black cross-hatch agents being compromised. The action set for agent $i$ is represented by the black lines to boxes, which are resources, i.e., the action set for agent $i$ is $\{\{r_i\}, \{r_6\}\}$. Each resource $r_j$ is given a value $v_j$ in the box. The welfare function is $W(a) = \sum_{r_j \in R(a)} v_j$, and each agent is endowed with the marginal contribution utility $MC_i$.}
        \label{fig:mc_ex_graph}
    \end{subfigure}
    \begin{subfigure}[b]{0.48\textwidth}
        \centering
        \includegraphics[scale=0.5]{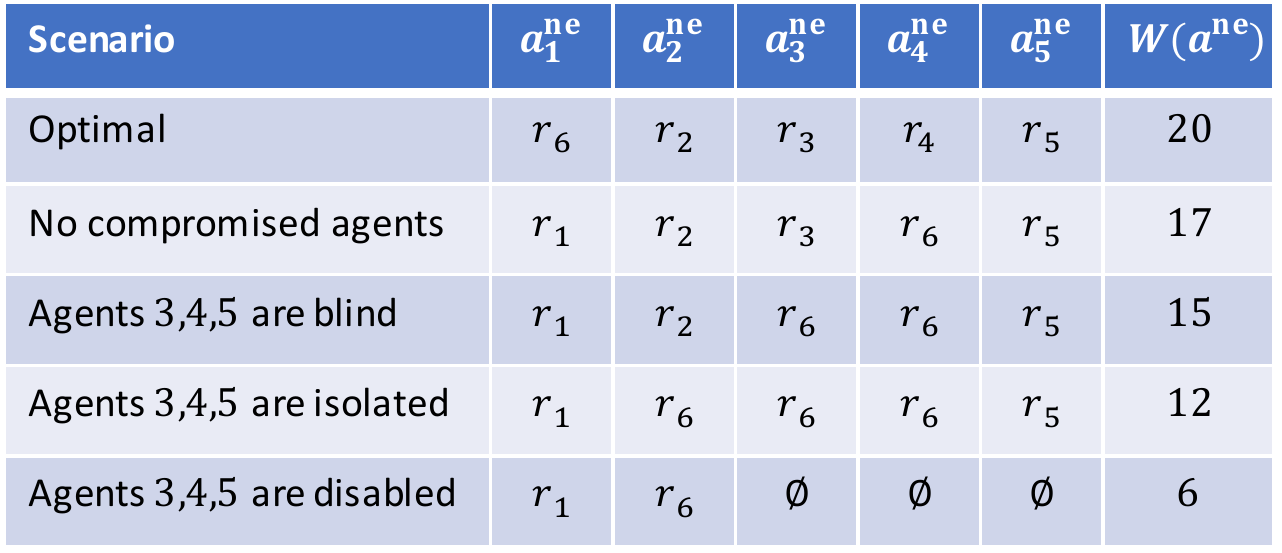}
        \caption{Worst-case Nash equilibria (and corresponding evaluations) for five scenarios. The first two assume that no agents have been compromised - note that the optimal allocation is also a NE. The final three assume that agents $3, 4, 5$ have been compromised in the same way.}
        \label{fig:mc_ex_table}
    \end{subfigure}
    \caption{A valid utility game where some agents have been compromised. We see that making agents $3,4,5$ blind causes agents $3$ and $4$ to both choose $r_6$. When $3,4,5$ are blind and $1,2$ cannot see their actions (i.e., $3,4,5$ are isolated), this additionally causes agent 2 to choose $r_6$. Finally, when $3,4,5$ are disabled, we see that they no longer contribute to the welfare of the system. While not every problem instance would degrade in this manner, we shall see in the results of this paper that this example is indicative of worst-case behavior. }
    \label{fig:mc_ex}
\end{figure}

We denote $K = B \cup I \cup D$ as the set of all compromised agents; i.e., $K$ may contain any combination of blind, isolated, and disabled agents.
Note that a blind or isolated agent can still select among its usual actions and its action choice still contributes to the system objective $W$ despite the denied observations.
We denote a VUG $(N, \aaa, \{U_i\}_i, W)$ with compromised agents as $G = (N, \aaa, \{U_i\}_i, W, (B, I, D))$. We also denote $\gee_k$ as the set of all such games where $|K| \leq k$. For an example of a VUG with compromised agents, see Figure \ref{fig:mc_ex}.

\subsection{Evaluating a Design: Price of Anarchy}

Throughout this paper, we focus on the solution concept of pure Nash equilibrium so as to abstract away the mechanics of specific learning rules and algorithms. We define agent $i$'s \emph{best response set} for an action profile $a_{-i}\in\aaa_{-i}$ as $B_i(a_{-i}) :=  \argmax_{a_i\in\aaa_i}\ U_i\left(a_i,a_{-i}\right)$.

An action profile $a^{\rm ne}\in\aaa$ is known as a \emph{pure Nash equilibrium} if for each agent $i$,
\vs
\begin{equation}
a^{\rm ne}_i \in B_i\left(a^{\rm ne}_{-i}\right). \label{eq:pnedef}
\vs
\end{equation}
That is, all agents are best-responding to each other.
The set of pure Nash equilibria of system $G$ is denoted ${\rm PNE}(G)$.

Accordingly, we measure the effectiveness of a given utility design by its \emph{price of anarchy}, comparing the quality of the Nash equilibria associated with the designed game against the optimal action profile for that multiagent problem:
\begin{equation}\label{eq:Qpess}
\poa\left(\gee_k\right)  \triangleq \inf_{G \in \mathcal{G}_k}
\frac{
	\min\limits_{a\in\PNE(G)}W(a)}
	{\max\limits_{a\in\aaa(G)}W(a)} \in [0,1],
\end{equation}
where $\PNE(G)$ denotes the set of pure Nash equilibria of problem $G$. Note that the closer the price of anarchy is to $1$, the more desirable the system-level performance.

\section{Results}

In this section, we demonstrate guarantees about the price of anarchy in a complex system when one or more agents in the system have been compromised. The effect this has on the price of anarchy depends somewhat on whether the agents in the compromised set $K$ are blind, isolated, or disabled.

\subsection{Effects of Denied Observations on Valid Utility Games}

\begin{theorem} \label{thm:vug 2+k}
    Let $G \in \gee_k$ be a valid utility game satisfying Definition~\ref{def:vug}, where agents in $K \subseteq N$ have been compromised with $|K|=k$. If at least one agent is disabled, then $\poa(\gee_k) = 0$. Otherwise,
    \begin{equation}
        \poa(\gee_k) = \max(1/(2 + |K|, 1/n))
    \end{equation}
\end{theorem}
\vspace{3mm}

Before giving the formal proof, we give a brief overview and some discussion of the significance of this result. It should be clear that having a disabled agent can be arbitrarily bad, thus the $\poa$ of 0 should not be surprising.
In order to show the remaining cases, we leverage the properties in Definition \ref{def:vug} and the definitions of blind and isolated agents to give a lower bound on $\poa(\gee_k)$. We then consider a subclass of VUGs where agents are endowed with a Shapley value utility function~\cite{Gopalakrishnan2010} as an example to show that the lower bound is tight.

Perhaps unintuitively, blind agents and isolated agents affect the $\poa$ in the same way; the information provided to the uncompromised agents by the actions of the blind agents has no effect.
The key deterrent to the $\poa$ is that the compromised agents do not consider the actions of others, not that others cannot see the actions of the compromised agents.

Another way to look at this is to think about a directed graph $(V, E)$, where each node represents an agent and an edge $(i, j)$ in the graph means that agent $j$'s utility function depends on the action of agent $i$. In a general sense, one might expect that the more ``connected" the graph, the better the resulting $\poa$. Under the nominal setting, where no agent is compromised, the graph is complete, and we have $\poa(\gee_0) = 1/2$.
When a single agent $i$ becomes blind, every edge $(j, i)$ for $j \neq i$ is removed from the graph, but all edges $(i, j)$ remain.
According to Theorem \ref{thm:vug 2+k}, this results in a decrease in the $\poa$ to $\poa(\gee_1)=1/3$.
If agent $i$ becomes isolated, this further removes all edges $(i, j)$ from the graph.
However, Theorem \ref{thm:vug 2+k} shows that the price of anarchy is unchanged at $\poa(\gee_1)=1/3$.

We note here that a similar result can be found in \cite{Grimsman2018a}, however the setting in \cite{Grimsman2018a} is different in that the solution concept is not that of NE: instead, the paper focuses on the allocation resulting from a sequential greedy algorithm. Additionally, \cite{Grimsman2018a} only considers the MC utility function, whereas in this work, we allow any utility function that satisfies properties 2) and 3) in Definition~\ref{def:vug}.

\begin{proof}
First we establish the case where agent $i \in K$ is disabled. Then one could construct an example with $W$ and $a_i \in \aaa_i$ such that $W(a_i, a_{-i})$ is arbitrarily large and $W(a_{-i})=0$ for any $a_{-i}$.
Since agent $i$ is forced to choose $\emptyset$, we see that $\poa(\gee_k) = 0$.

For the remainder of the proof we assume that all agents in $K$ are either blind or isolated. In the first case we assume that $|K| < n-1$. We show through the properties of Definition \ref{def:vug} and the definition of blind and isolated agents that $1/(2 + |K|)$ is a lower bound on $\poa(\gee_k)$. Then we show that the bound is tight by choosing a particular welfare function $W$ and utility $U$ such that $W(a^{\rm ne}) / W(a^\opt) = 1/(2 + |K|)$.

To see the lower bound, let $a^{\rm opt}$ be an optimal allocation, i.e., $a^{\rm opt}$ is in the $\argmax$ of the denominator in \eqref{eq:Qpess}. We also denote $a_{j < i}$ to mean $a_1, \dots a_{i-1}$. Finally, denote $P_i = N \setminus (I \cup D \cup \{i\})$, i.e., $P_i$ is the set of agents whose actions agent $i \notin K$ ``observes". Then we see that
\begin{align}
    W(& a^{\rm opt})  \le  W(a^\opt, a^{\rm ne}) \label{eq:genlb1} \\
    \le & W(a^{\rm ne}) + \sum_i W(a^\opt_i, a^\opt_{j<i}, a^{\rm ne}) - W(a^\opt_{j<i}, a^{\rm ne}) \label{eq:genlb2} \\ 
    \le & W(a^{\rm ne}) + \sum_i W(a^\opt_i, a^{\rm ne}_{P_i}) - W(a^{\rm ne}_{P_i}) \label{eq:genlb3} \\ 
    \le & W(a^{\rm ne}) + \sum_{i \notin K} U_i(a^\opt_i, a^{\rm ne}_{P_i}) + \sum_{i \in K} W(a_i^\opt) \label{eq:genlb4} \\
    \le & W(a^{\rm ne}) + \sum_{i \notin K} U_i(a^{\rm ne}_i, a^{\rm ne}_{P_i}) + \sum_{i \in K} \tilde{U}_i(a_i^\opt) \label{eq:genlb5} \\
    \le & W(a^{\rm ne}) + 
    W(a^{\rm ne}) + \sum_{i \in K} \tilde{U}_i(a_i^{\rm ne}) \label{eq:genlb6} \\
    \le & W(a^{\rm ne}) + 
    W(a^{\rm ne}) + \sum_{i \in K} W(a_i^{\rm ne}) \label{eq:genlb7} \\
    \le & (2 + |K|) W(a^{\rm ne}), \label{eq:genlb8}
\end{align}
where \eqref{eq:genlb1} is true since $W$ is nondecreasing; \eqref{eq:genlb2} is true via telescoping; \eqref{eq:genlb3} is true by submodularity of $W$; \eqref{eq:genlb4} holds since the original $U_i$ satisfy 2) in Definition \ref{def:vug} (2nd term), and by submodularity of $W$ (3rd term); \eqref{eq:genlb5} is true by defintion of NE (2nd term) and by the utlities of the blind and isolated agents (3rd term); \eqref{eq:genlb6} is true since the original $U_i$ satisfy 3) in Definition \ref{def:vug} (2nd term) and by defintion of NE (3rd term); \eqref{eq:genlb7} is true by the defintion of $\tilde{U}_i$ for agents in $K$; and \eqref{eq:genlb8} is true since $W$ is nondecreasing.

To see the upper bound, consider a scenario where $W$ is of the form
\begin{equation} \label{eq:sepw}
    W = \sum_{r \in \mathcal{R}} W_r(|a|_r), 
\end{equation}
where $|a|_r$ denotes the number of agents which have selected resource $r$ under allocation $a$.
The functions $W_r: \{1, \dots, N\} \to \mathbb{R}$ are nonnegative, i.e., $W_r(i) \ge 0$; nondecreasing, i.e., $W_r(i+1) \geq W_r(i)$; and have decreasing marginal returns, i.e., $W_r(i+1) - W_r(i) \ge W_r(i+2) - W_r(i+1)$. When $W$ has this form, this represents a well-studied set of games called distributed resource allocation games (see, for instance \cite{hespanha2017noncooperative}).

We also assume that, before agents are compromised, all are endowed with the equal share ($\es$) utility, wherein each agent chooses an action according to the following:
\begin{equation}
    \es_i(a_i, a_{-i}) = \sum_{r \in R(a_i)} \frac{1}{|a|_r} W_r(|a|_r),
\end{equation}
i.e., when multiple agents choose the same resource, the agents divide the utility $W_r(|a|_r)$ equally. We note that the $\es$ utility is an instance of the more general Shapley value utility, also a subject of much study within the literature. 

Based on the construction of $W_r$ and therefore $W$, it should be clear that $W$ satisfies 1) in Definition \ref{def:vug}. Likewise, it should be immediately clear that when $\es$ is employed, 3) in Definition \ref{def:vug} is satisfied with equality. We can also see that 2) is satisfied since
\begin{align}
    W(a) - W(\emptyset, a_{-i}) = & \sum_{r \in R(a_i)} \frac{W_r(|a|_r)}{|a|_r}  - \frac{W_r(|a_{-i}|_r)}{|a_{-i}|_r}, \\
    \leq & \sum_{r \in R(a_i)} \frac{W_r(|a|_r)}{|a|_r} = \es_i(a).
\end{align}
Therefore, $G = (N, \aaa, \{\es_i\}_i, W)$, where $W$ has the form in \eqref{eq:sepw} is a VUG.

Assume that the example shown in Figure \ref{fig:k_blind} is such a game, where now a subset of agents $K$ are all blind. The blind agents are represented by the black cross-hatch circles, and the other agents as blue circles. Each agent has access to its own resource (the box closest to it) and a central resource. The value $v_r$ of resource $r$ is the value listed in the box, where $\varepsilon, \delta > 0$ are small. Then $W_r = v_r$, i.e., $W(a) = \sum_{r \in R(a)} v_r$. We see that the agents in $K$ will all choose the central resource, since they all act independently. We also see that in any NE, all agents not in $K$ are also incentivized to choose the central resource, since they are endowed with ES. Therefore, $W(a^{\rm ne}) = 1$. The optimal allocation is for one agent not in $K$ to choose the central resource and the remaining agents to choose their alternates, implying that $W(a^\opt) = 1 + (n - |K| - 1)(1/n - \delta) + |K|(1-\varepsilon)$. As $\varepsilon, \delta \to 0$, and $n \to \infty$, we see that
\begin{equation}
    \frac{W(a^{\rm ne})}{W(a^\opt)} \to \frac{1}{2 + |K|}.
\end{equation}

\begin{figure}
    \centering
    \includegraphics[width=0.4\textwidth]{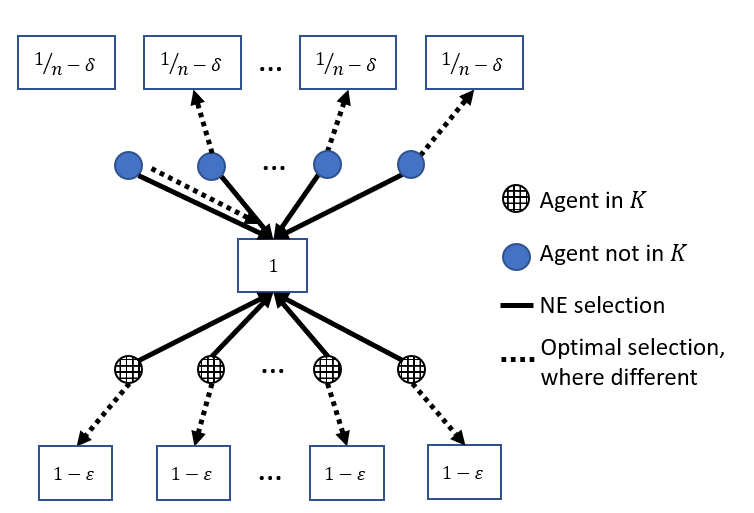}
    \caption{\small An example game used in the proof for Theorem \ref{thm:vug 2+k}, a VUG where now a subset of agents $K$ are all blind. Each agent has access to its own resource (the box closest to it) and a central resource. The value $v_r$ of resource $r$ is the value listed in the box, where $\varepsilon, \delta > 0$ are small, and $W(a) = \sum_{r \in R(a)} v_r$. We see that the agents in $K$ will all choose the central resource, since they all act independently. Agents not in $K$ are endowed with ES, and therefore are also incentivized to choose the central resource. This implies $W(a^{\rm ne}) = 1$ and as $\varepsilon, \delta \to 0$ and $n \to \infty$, $W(a^\opt) \to 2+|K|$. }
    \label{fig:k_blind}
\end{figure}

\end{proof}

\subsection{Marginal Contribution Utility}

In this section, we consider the use of the marginal contribution utility and whether using this specific utility function design can offset the decrease in $\poa$ that we saw in Theorem \ref{thm:vug 2+k}. We show that marginal contribution utility can give a higher $\poa$ in the presence of blind agents. In order to do so, we define $\gee_k^\mc \subseteq \gee_k$ as the subset of games in $\gee_k$ which leverage $\mc_i$ for all agents.

\begin{theorem} \label{thm:mc good}
Let $G\in \gee_k^\mc$ be any valid utility game which uses the marginal contribution utility, where agents $K \subseteq N$ are compromised. If one agent is disabled, then $\poa(\gee_k^\mc) = 0$. Otherwise
\begin{equation}
        \poa(\gee_k^\mc) = \left\{ 
        \begin{array}{cl}
            \frac{1}{1 + |K|}, & \text{if $|B| > 0$,} \\
            \max\left(\frac{1}{2 + |K|}, \frac{1}{n}\right), & \text{if $|B| = 0$.} \\
        \end{array}
        \right.
    \end{equation}
\end{theorem}

\begin{figure}
    \centering
    \includegraphics[scale=0.6]{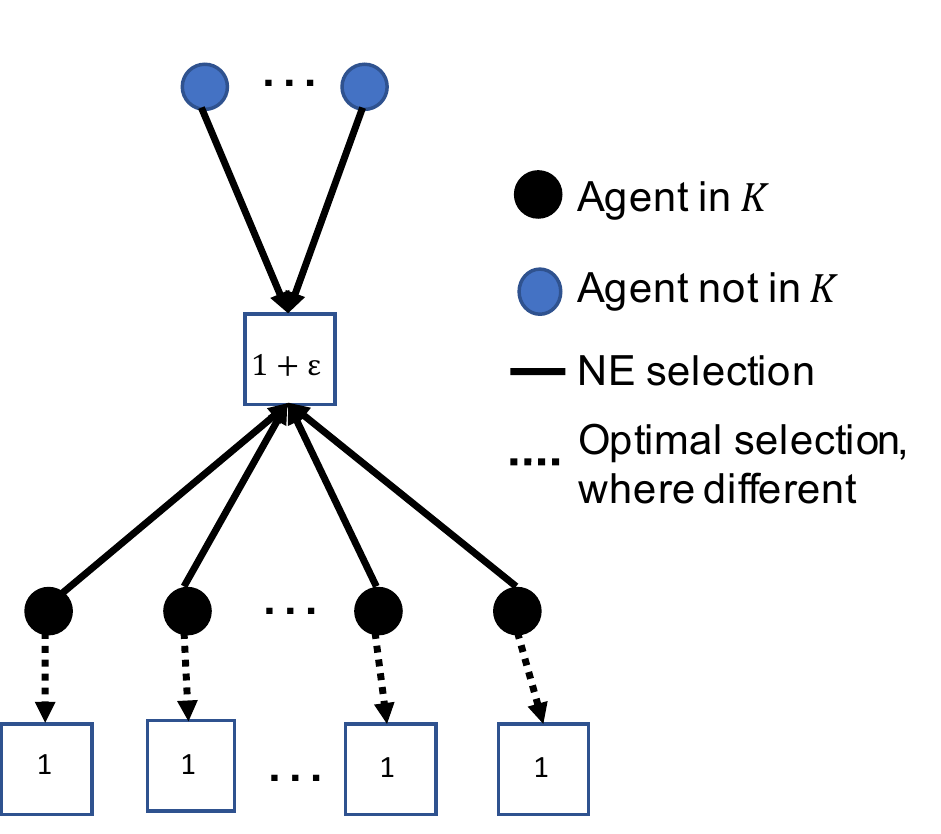}
    \caption{\small An example game used in the proof for Theorem \ref{thm:mc good}. As with the other figures, compromised agents are the black cross-hatch circles and the agents not in $K$ are the blue circles. Each resource $r$ has a value $v_r$, where $W(a) = \sum_{r \in R(a} v_r$. A NE selection yields $W(a^{\rm ne}) = 1 + \varepsilon$, since the blue agents have no other available resources, and the agents in $K$ act independently. The optimal selection is for the agents in $K$ to choose their alternate resource, i.e., $W(a^\opt) = |K| + 1 + \varepsilon$. Therefore, as $\varepsilon \to 0$, $W(a^\opt) \to 1 + |K|$.}
    \label{fig:mc_blind}
\end{figure}

\begin{proof}
    We note that the values for $\poa(\gee_k)$ shown in Theorem \ref{thm:vug 2+k} are lower bounds for $\poa(\gee_k^\mc)$. Therefore, we need only show that having one blind agent increases the lower bound for $\poa(\gee_k^\mc)$, then show that all these lower bounds are tight.

    As with Theorem \ref{thm:vug 2+k}, we use submodularity and monotonicity to show that when $|B| > 0$, $\poa(\gee_k^\mc) \geq \frac{1}{1+|K|}$, and then proceed to provide a canonical example to show that this lower bound on $\poa$ is tight. As a matter of notation, for $a, b \in \mathcal{A}$, we denote $W(a, b)$ to mean $W(c)$, where $c = \{a_i \cup b_i\}_i$. Futhermore, for some $J \subseteq N$, we use $a_J$ to mean $\{a_i\}_{i \in J}$.

    To see the lower bound, suppose that $B \neq \emptyset$ and consider the following:
    \begin{align}
        W(a^{\rm opt}) \leq & W\left(a^{\rm opt}, a^{\rm ne}_{B}\right), \label{eq:mcpoalb1} \\
        \leq & W\left( a^{\rm opt}_{N \setminus K}, a^{\rm ne}_{B}\right)  + \sum_{i \in K} W \left(a_i^{\rm opt} \right), \label{eq:mcpoalb2} \\
        = & W\left( a^{\rm opt}_{N \setminus K}, a^{\rm ne}_B\right)  + \sum_{i \in K} W \left(a_i^{\rm ne} \right), \label{eq:mcpoalb3} \\
        \leq & W\left( a^{\rm opt}_{N \setminus K}, a^{\rm ne}_B\right) + W(a^{\rm ne}_B) + (|K|-1) W(a^{\rm ne}) \label{eq:mcpoalb4}
    \end{align}
    where \eqref{eq:mcpoalb1} is true by monotonicity, \eqref{eq:mcpoalb2} is true by submodularity, \eqref{eq:mcpoalb3} is true since agents in $K$ are either blind or isolated, and \eqref{eq:mcpoalb4} is true by monotonicity.
    
    Once an agent in $K$ has chosen an action, that agent has no incentive to deviate, regardless of how the other agents behave. Therefore, we can consider a ``sub game" ${\bar{G} = (N \setminus K, \bar{\mathcal{A}}, \overline{\mc}, \overline{W})}$ among only the non-compromised agents, assuming that the blind and isolated agents have made their choices. In this sub game, the non-compromised agents seek to maximize the welfare function ${\overline{W} : \bar{\mathcal{A}}_i \to \mathbb{R}}$, where $\bar{A}:= \Pi_{i \notin K}\mathcal{A}_i$, such that
    \begin{equation}
        \overline{W}(\bar{a}) = W(\bar{a}, a^{\rm ne}_B) - W(a^{\rm ne}_B), 
    \end{equation}
    for $\bar{a} \in \bar{\mathcal{A}}_i$. Note that $\overline{W}$ is also submodular monotone, with $\overline{W}(\emptyset) = 0$. The agents are endowed with the following utility function
    \begin{equation}
        \overline{\mc}_i(\bar{a}_i, \bar{a}_{-i}) = \overline{W}(\bar{a}_i, \bar{a}_{-i}) - \overline{W}(\bar{a}_{-i}).
    \end{equation}
    It can be easily shown that is a VUD. Therefore, we know from \cite{Vetta2002} that $2\overline{W}(\bar{a}^{\rm ne}) \geq \overline{W}(\bar{a}^{\rm opt})$, where ${\bar{a}^{\rm opt} \in \argmax_{\bar{a}} \overline{W}(a)}$. It is also important to note that by design, $\bar{a}^{\rm ne}$ is also a NE profile of actions for agents not in $K$ for the \emph{original} game $G$, assuming that agents in $B$ choose $a^{\rm ne}_B$.
    
    Returning to \eqref{eq:mcpoalb4}, we see that
    \begin{align}
        W(a^{\rm opt}) & \leq W\left( a^{\rm opt}_{N \setminus K}, a^{\rm ne}_B\right) - W(a^{\rm ne}_B) \nonumber\\
        &\ \ \ + 2W(a^{\rm ne}_B) + (|K| - 1)W(a^{\rm ne}) \label{eq:mcpoalb5} \\
        =& \overline{W}\left( a^{\rm opt}_{N \setminus K} \right) + 2W(a^{\rm ne}_B) + (|K| - 1)W(a^{\rm ne}) \label{eq:mcpoalb6} \\
        \leq & \overline{W}\left( \bar{a}^{\rm opt} \right) + 2W(a^{\rm ne}_B) + (|K| - 1)W(a^{\rm ne}) \label{eq:mcpoalb7} \\
        \leq & 2\overline{W}\left( \bar{a}^{\rm ne} \right) + 2W(a^{\rm ne}_B) + (|K| - 1)W(a^{\rm ne}) \label{eq:mcpoalb8}\\
        =& 2W\left( a^{\rm ne}_{N \setminus K}, a^{\rm ne}_B\right) + (|K| - 1)W(a^{\rm ne}_K) \label{eq:mcpoalb9} \\
        \leq & (1 + |K|) W(a^{\rm ne}), \label{eq:mcpoalb10}
    \end{align}
    where \eqref{eq:mcpoalb5} is trivially true, \eqref{eq:mcpoalb6} is true by defintion of $\overline{W}$, \eqref{eq:mcpoalb7} is true by definition of $\bar{a}^{\rm opt}$, \eqref{eq:mcpoalb8} is true since $\bar{G}$ is a VUG, \eqref{eq:mcpoalb9} is true by definition of $\overline{W}$, and \eqref{eq:mcpoalb10} is true by monotonicity. Thus for any $G$ that meets the requirements of the theorem statement, and for any $a^{\rm ne}$, it follows that
    \begin{equation}
        \frac{W(a^{\rm ne})}{W(a^{\rm opt})} \geq \frac{1}{1 + |K|},
    \end{equation}
    implying that this is also a lower bound for $\poa(\mc, K)$.

\begin{figure}
    \centering
    \includegraphics[scale=0.6]{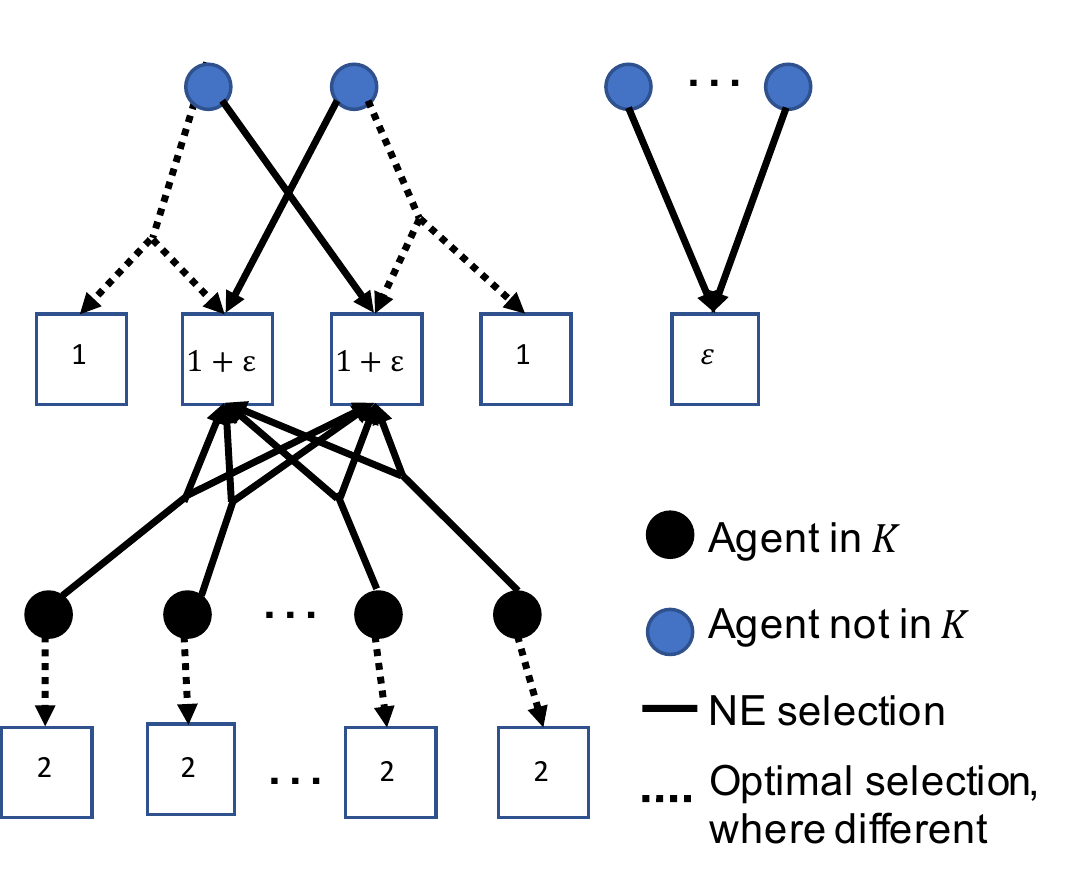}
    \caption{\small An example game used in the proof for Theorem \ref{thm:mc good}, for the case where $K$ is only isolated agents. We use the same $W$ and notation as Figure \ref{fig:mc_blind}, except with different resource values and action sets. As with previous examples, agents in $K$ act independently, choosing the two $1 + \varepsilon$ resources. The agents not in $K$ are not informed of these decisions, so a worst-case NE is where the two agents which also have access to these two resources also select them. The remaining agents not in $K$ have only access to one $\varepsilon$ resource. Therefore, in this case, $W(a^{\rm ne}) = 2 + 3\varepsilon$. The optimal allocation, on the other hand, is for all resources to be selected. Thus as $\varepsilon \to 0$, $W(a^{\rm ne})/W(a^\opt) \to 1/(2 + |K|)$.}
    \label{fig:mc_noblind}
\end{figure}

We now show that the lower bounds established above are tight. First consider the problem instance in Figure~\ref{fig:mc_blind}. Again, compromised agents are the black cross-hatch circles and the agents not in $K$ are the blue circles. As with the other examples in this paper, each resource $r$ has a value $v_r$, where $W(a) = \sum_{r \in R(a} v_r$. It should be clear that a NE selection yields $W(a^{\rm ne}) = 1 + \varepsilon$, since the blue agents have no other available resources, and the agents in $K$ act independently. The optimal selection is for the agents in $K$ to choose their alternate resource, i.e., $W(a^\opt) = |K| + 1 + \varepsilon$. Therefore, as $\varepsilon \to 0$, we see that
\begin{equation}
    \frac{W(a^{\rm ne})}{W(a^\opt)} \to \frac{1}{1 + |K|}.
\end{equation}
Note that this holds for any combination of isolated and blind agents in $K$, and as long as $|K| < n$.

In the case where $K = N$, then in the example in Figure \ref{fig:mc_blind} one agent will still choose the $1 + \varepsilon$ resource. Here we see that as $\varepsilon \to 0$, $W(a^{\rm ne})/W(a^\opt) \to 1/n$.

In the case where $|K| < n-1$ and there are no blind agents in $K$, we invoke the example in Figure \ref{fig:mc_noblind}, which uses the same $W$ and notation as Figure \ref{fig:mc_blind}, except with different resource values and action sets. As with previous examples, agents in $K$ act independently, choosing the two $1 + \varepsilon$ resources. The agents not in $K$ are not informed of these decisions, so a worst-case NE is where the two agents which also have access to these two resources also select them. The remaining agents not in $K$ have only access to one $\varepsilon$ resource. Therefore, in this case, $W(a^{\rm ne}) = 2 + 3\varepsilon$. The optimal allocation, on the other hand, is for all resources to be selected. Thus as $\varepsilon \to 0$,
\begin{equation}
    \frac{W(a^{\rm ne})}{W(a^\opt)} \to \frac{1}{2 + |K|}.
\end{equation}
\end{proof}

\section{Simulation} \label{sec:sim}

 In this section we present empirical findings from the results of running a simulation of stochastic learning dynamics applied to a VUG in which agents are endowed with a marginal contribution utility function.
 We simulate the popular \emph{log-linear learning} dynamics~\cite{Alos-Ferrer2010,Marden2012a} to validate the results and explore the effect of ``noisy'' behavior on these low-quality equilibria.
Log-linear learning operates in discrete steps at times $t = 0,1,\ldots,$ producing a sequence of joint actions $a(0),a(1),\ldots.$
We assume agents begin with an arbitrary joint action $a(0)\in\mathcal{A}$, and let $a(t) = (a_i,a_{-i})\in\mathcal{A}.$
At time $t\in\mathbb{N}$, agent $i\in N$ is selected uniformly at random to update its action for time $t+1$; all other agents' actions will remain fixed for time $t+1$.
At time $t+1$, agent $i$ chooses action $a_i\in\aaa_i$ with probability
\begin{align}
p_i^{a_i}(t+1) = \frac{e^{U_i(a_i,a_{-i}(t))/T}}{\sum_{\tilde{a}_i\in\aaa_i}e^{U_i(\tilde{a}_i,a_{-i}(t))/T}}.\label{e:LLL dynamics}
\end{align} 
``Temperature'' parameter $T>0$ dictates an updating agent's degree of rationality and is identical for all agents $i\in N$. 
As $T\to 0$, agents are increasingly likely to select utility-maximizing actions, and as $T\to \infty$, agents tend to choose their next actions uniformly at random.
The joint action at time $t+1$ is $a(t+1) = (a_i(t+1),a_{-i}(t)).$

\begin{figure}
    \centering
    \begin{subfigure}[b]{0.48\textwidth}
        \centering
        \includegraphics[scale=0.45]{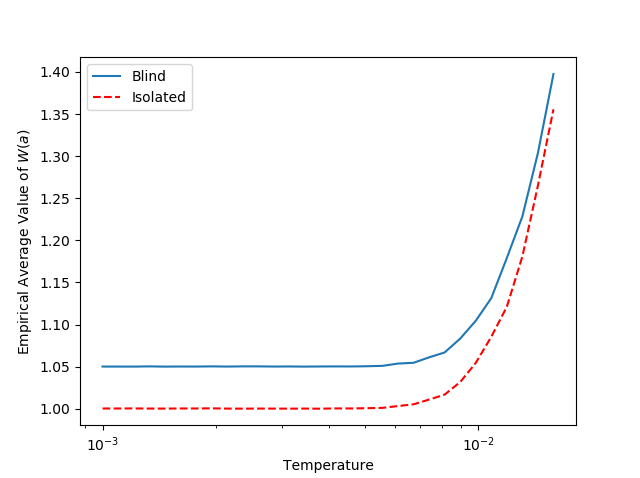}
        \caption{Simulations with temperatures ranging from $0.001$ to $0.016$.}
        \label{fig:blind v isolated}
    \end{subfigure}
    \begin{subfigure}[b]{0.48\textwidth}
        \centering
        \includegraphics[scale=0.45]{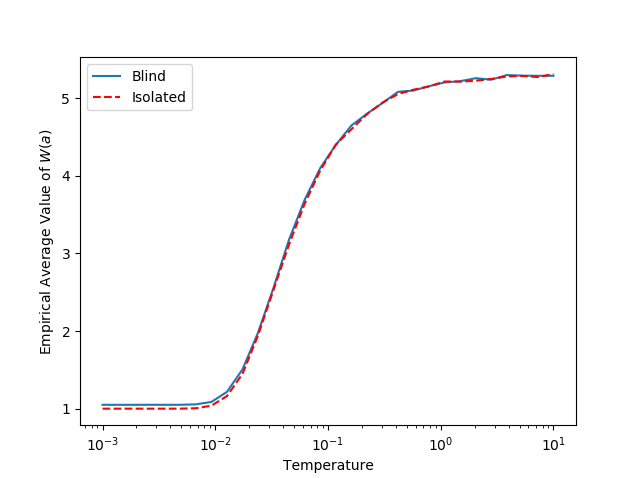}
        \caption{Simulations with temperatures ranging from $0.001$ to $10$.}
        \label{fig:blind v isolated macro}
    \end{subfigure}
    \caption{\small Plots representing simulation results of the game shown in Figure \ref{fig:k_blind}, with $n=10$, $|K|=9$, with the exceptions that the agent not in $K$ has the MC utility and its alternate resource has value $\varepsilon=0.05$. For each trace, log-linear learning is run for 200,000 iterations for each temperature value.
    The solid blue trace corresponds to all agents in $K$ being blind, and the dashed red trace corresponds to all agents in $K$ being isolated.
    Note that for very low temperatures (effectively the agents are playing asynchronous best-response dynamics), blindness has a slight advantage over isolation in accordance with Theorem~\ref{thm:mc good}.}
    \label{fig:my_label}
    \vspace{-5mm}
\end{figure}

 We run log-linear learning on the game depicted in Figure~\ref{fig:k_blind} with the following exceptions: there is only $1$ agent not in $K$, it is endowed with the MC utility, and its ``alternate" resource option has value $\varepsilon$. We use $n=10$, $|K|=9$, and $\varepsilon=0.05$, and for each value of $T$, we report the average value of $W(a)$ for 200,000 trials. The difference in a game in which all the compromised agents are isolated versus one in which at least one of these agents is blind can be seen in Figure \ref{fig:blind v isolated}.  
 The compromised agents in $K$ are all blind in the first simulation (solid blue), and all are isolated in the second (dashed red). 
 
 For this game, since $\varepsilon=0.05$, the optimal selection of resources yields a value of the welfare function of $9.55$. The Nash equilibrium yields a value of 1 when all agents in $K$ are isolated, and $1.05$ when at least one agent in $K$ is blind. In the simulation, the game in which all the agents were blind had a minimum average value of the welfare function of $1.050015$, hence the price of anarchy is $0.10995$. The game in which all the agents were isolated had an minimum average value of welfare function of $1.000034$, giving a price of anarchy of $0.1047$. These values of the price of anarchy differ slightly from those given by Theorem~\ref{thm:mc good} since $\varepsilon \ne 0$. As temperature increases, the instability of the Nash equilibrium becomes apparent, and the average of the welfare function increases with an increase in temperature until this value is indistinguishable from that produced by a purely random strategy as seen in Figure \ref{fig:blind v isolated macro}.\par
 
 An intriguing aspect of this example is that the Nash equilibrium, representing a worst-possible Nash equilibrium in the class of games $\gee^{\rm MC}_9$, is actually among the worst action profiles in the game.
 Hence, a large value of $T$ (i.e., agents selecting actions uniformly at random) results in play that is of far higher quality than the Nash equilibrium.

\section{Conclusion}

This paper presents new results regarding the robustness of valid utility games to lost or denied information transfer among agents, and illustrates that in a broad array of submodular maximization games, performance guarantees degrade gracefully with the degree of information denial.
Our results also show a variety of intriguing phenomena, such as the notion that isolation is no worse than blindness in worst case and that marginal-contribution utility functions provably outperform the baseline performance guarantees.
Many open questions remain, including that of which utility functions are \emph{optimal} in this context and how arbitrary unstructured information losses impact performance guarantees.

\bibliographystyle{ieeetr}
\bibliography{library/library}

\end{document}

%% file: philipMacros.tex
\newcommand{\mc}{{\rm MC}}

\newcommand{\arr}{\mathbb{R}}

\newcommand{\PNE}{{\rm PNE}}

\newcommand{\vs}{\vspace{-1mm}}

\newcommand{\aaa}{\mathcal{A}}

\DeclareMathOperator*{\argmax}{arg\,max}

\newcommand{\gee}{\mathcal{G}}

\newcommand{\poa}{\ensuremath{{\rm PoA}}}

